# POST-HOC INTERPRETABILITY BASED PARAMETER SELECTION FOR DATA ORIENTED NUCLEAR REACTOR ACCIDENT DIAGNOSIS SYSTEM

**Chengyuan Li**
Science and Technology on Reactor System Design Technology Laboratory, Nuclear Power Institute of China, Chengdu, China

**Meifu Li**
Science and Technology on Reactor System Design Technology Laboratory, Nuclear Power Institute of China, Chengdu, China

**Zhifang Qiu**
Science and Technology on Reactor System Design Technology Laboratory, Nuclear Power Institute of China, Chengdu, China

**ABSTRACT**

During applying data-oriented diagnosis systems to distinguishing the type of and evaluating the severity of nuclear power plant initial events, it is of vital importance to decide which parameters to be used as the system input. However, although several diagnosis systems have already achieved acceptable performance in diagnosis precision and speed, hardly have the researchers discussed the method of monitoring point choosing and its layout. For this reason, redundant measuring data are used to train the diagnostic model, leading to high uncertainty of the classification, extra training time consumption, and higher probability of overfitting while training. In this study, a method of choosing thermal hydraulics parameters of a nuclear power plant is proposed, using the theory of post-hoc interpretability theory in deep learning. At the start, a novel Time-sequential Residual Convolutional Neural Network (TRES-CNN) diagnosis model is introduced to identify the position and hydrodynamic diameter of breaks in LOCA, using 38 parameters manually chosen on HPR1000 empirically. Afterwards, post-hoc interpretability methods are applied to evaluate the attributions of diagnosis model's outputs, deciding which 15 parameters to be more decisive in diagnosing LOCA details. The results show that the TRES-CNN based diagnostic model successfully predicts the position and size of breaks in LOCA via selected 15 parameters of HPR1000, with 25% of time consumption while training the model compared the process using total 38 parameters. In addition, the relative diagnostic accuracy error is within 1.5% compared with the model using parameters chosen empirically, which can be regarded as the same amount of diagnostic reliability.

## 1. INTRODUCTION

Recognition of the initiating event and analysis of the severity of the accident is critically important in the early stages of a reactor accident, as this will directly influence the actions taken by the operator to prevent further worsening of the accident. If the operator adopts the wrong course of action, the risk of a radioactive release is increased. As the reactor is a non-linear system of high degree of complexity and the instrument measurement system has a large number of detection parameters available to the operator, it is a challenging task to assist the operator in making an accurate and timely diagnosis of the accident.

In the attempt to improve the efficiency of diagnosing reactor accidents, researchers have developed diagnostic models based on a diverse range of principles[1]. With the rise of deep learning algorithms, this approach has a better performance in the field of pattern recognition. As the task of diagnosing reactor accidents can be considered as a type of pattern recognition task, diagnostic methods using deep learning tend to have outstanding capabilities in data-based diagnostic methods. However, there are several shortcomings in the work done by previous researchers.

Firstly, in terms of model construction work, a variety of accident diagnosis models have been developed using deep learning. In recent years, an accident diagnosis model based on LSTM and CNN has been proposed by Saeed et al., which can not only identify the 'unknown' transient type, but also calculate the severity of the accident[2]. Wang et al. used convolution layer instead of full connection layer in traditional GRU neural network, whose super-parameters are searched by particle swarm optimization algorithm, which achieved good results in experiments[3]. Nevertheless, they only illustrate the superiority of the models in terms of their diagnostic accuracy, without explaining why they perform well on the diagnostic task in terms of the models themselves.

Secondly, as a large amount of data is generated during reactor accident transients, a series of methods have been proposed previously to reduce the dimensionality of the monitoring data in order to speed up the convergence of the training of the model. Among them, the PCA linear dimensionality reduction method is the most dominant technique of choice[4–7]. Although PCA can reduce the input data to the model, researchers have not shown the reasons why the dimensionality reduction is valid. It is notable that the validity mentioned here does not refer to the fidelity of the data.

Thirdly, the diagnostic model requires several monitored parameters of the accident transient process to be selected as input to the model. However, there has not yet been a methodology for the selection of parameters and the choice of parameters is more of an empirical approach. In addition, as the final decision on the diagnostic result has to be made by the operator rather than the diagnostic system, the diagnostic system needs to provide not only the diagnostic result but also effective detection parameters to assist the operator in the manual diagnosis. Research work in this area is largely absent.

This paper therefore offers two points of work in the area of assisting the operator in accident diagnosis. The first is the development of a new accident diagnosis model. The secondary point is to propose a methodology that provides more significant monitoring parameters for accident diagnosis results, inspired by the post-hoc interpretability theory in deep learning. The contributions of this paper are listed as follows:

1) A model structure called Time-sequential Residual Convolutional Neural Network (TRES-CNN) is proposed for time series data, effectively avoiding overfitting, and multiple outputs for category and regression labels.
2) An integrated deep learning interpretability approach is proposed to implement parameter importance analysis, which improves the validity of the input data and can be used by operators as a basis for artificially diagnosing the accident.
3) The proposed model outperforms the baseline in terms of relative error in identifying the size and location of ruptures. Further analysis demonstrates the effectiveness of the proposed method for accident diagnosis.

The whole paper is organized as follows. We describe our proposed diagnostic model in Section 2. In Section 3, we introduce the experiments and conduct an analysis and discussion. Finally in section 4 we conclude the paper and discuss future work.

## 2. PROPOSED METHODS

### 2.1 Construction of the Diagnostic Model

Convolution Neural Network (CNN) is a kind of conventional deep neural network structure, which can automatically extract the abstract features of original data via its deep structure. Compared with Fully Connected Neural Networks (FCNN), CNN is generally thought to be more adaptable to massive data set, due to its pooling mechanism to down sample data while extracting abstract features. Therefore, CNN is more suitable than other FCNNs to learn NPP's transient data, which is high dimensional, highly nonlinear and massive.

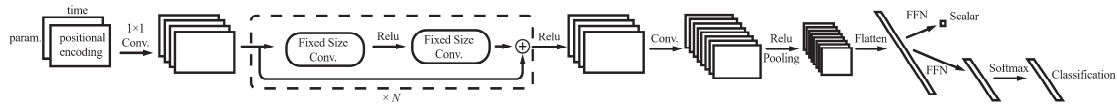

**Figure 1 Schematic diagram of TRES-CNN**

However, preliminary CNN based pattern recognition model cannot learn the sequential features of the time serial data intrinsically. Inspired by the Transformer model[8], which is the state-of-the-art framework used in time sequential tasks, a positioning encoding mechanism is used to extend original models support for time serial data. The positioning encoding method is to define certain values which are attached to each element and each dimension of any sampled data, and the encoding function is defined as:

$$PE_{(pos, 2i)} = \sin\left(pos / 10000^{2i/dim}\right)$$
$$PE_{(pos, 2i+1)} = \cos\left(pos / 10000^{2i/dim}\right) \quad (1)$$

where $pos$ is the position of the sampled data, $i$ is the position of the parameter and dim is the total parameters collected from the simulation results. A map of the position coding is shown below as Figure 2, with the horizontal axis showing the coding corresponding to the different parameters at each sampling moment, and the vertical axis showing the coded information of the parameters on the time axis.

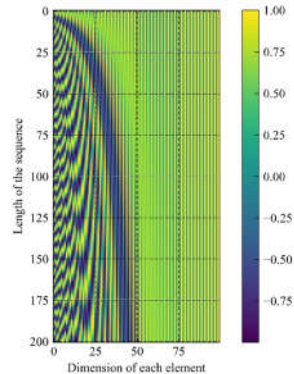

**Figure 2 Positional encoding map**

Moreover, in order to prohibit the model from forgetting the time sequential information while multiplying convolutional layers, which is to incrementally learn the deep features of the data, several residual blocks are applied to the diagnostic model inspired by ResNet[9]. The schematic structure of a residual block is shown in Figure 3.

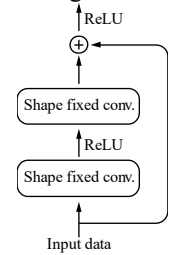

**Figure 3 A single residual block**

After extracting the features of the data from each transient, the diagnostic layers are needed to predict the position and size of the break. Identifying position is a classification problem, and evaluating the size is a regression problem, which need feed-forward network layer and Softmax layer respectively. Ultimately, the general framework of the TRES-CNN is constructed, and the schematic diagram of TRES-CNN is shown in Figure 1.

Compared with the sequential oriented deep learning models (e.g., RNN, GRU and LSTM), the proposed TRES-CNN

is not only suitable for tasks with time series, but also inherits the parallel computing capability of CNN. In this regard, TRES-CNN is a promising method for diagnostic model of LOCA.

## 2.2 Parameter Significance Analysis Method

However, the surge in performance of DL based diagnostic models is often achieved by increasing model complexity, which turns such models into "black box" pathways and creates uncertainty about how they operate and how they will ultimately decide. Previously, researchers argued that the interpretability of diagnostic models using AI theories worth further study[10]. In this study, the methods to improve the interpretability of the diagnostic model are applied to figure out which parameters selected initially are worth more attention when diagnosing LOCAs.

Post-hoc interpretability means not building an interpretable learning algorithm at the outset, but rather building a black-box model such as a neural network and then back-propagating how the model discriminates[11]. Post-hoc interpretable methods are broadly classified into five categories, namely activation maximization-based, gradient-based, class activation mapping method-based, LIME-based and knowledge distillation-based methods. Among them, the gradient-based and LIME-based methods are able to directly calculate the importance of the input data on the output results. Considering that the gradient-based weighted class activation mapping (Grad-CAM++) has better applicability to CNN-based models than the conventional gradient-based approaches, and that the LIME approach constructs a linear model with local alternatives, which has a completely different mechanism from the gradient-based approach, a combination of the two can be used to better effect. Therefore, an integration model based on Grad-CAM++ and LIME is proposed as a means of evaluating the magnitude of the contribution of the input parameters to the output prediction of the LOCA diagnostic model.

Grad-CAM++ is a Post-hoc interpretable method derived from an improvement of Grand-CAM, which is able to give a more complete heat map of the model's attention compared to the original method and retains the advantages of the original method, i.e. being more applicable to models with CNN as the architecture and the fact that no modifications to the diagnostic model are required during the interpretation process. The idea of this method is to select the node with the highest Softmax (corresponding to the class with the highest confidence) to back-propagate the gradient of the last layer of the convolutional layer, and the average of the gradients of each feature map is used as the weight of that feature map. It is based on the basic assumption that the last layer of the convolutional neural network has sufficient abstraction information about the original data[12]. Since the prediction of break size in LOCA accident is a regression problem completed by CNN model, Grad-CAM++ is used as a parameter importance analysis tool for break size.

On the other hand, LIME is an interpretable method suitable for any complex classifier. Its overall goal is to determine an interpretable model on an interpretable representation that is a local fit to the classifier. It uses perturbations local to the sample point to analyze the impact of the model's output and retrains a simple model that fits well around the sample point for a single sample, which is usually a linear or decision tree model, and in turn obtains a significant attribution analysis of the model local to the sample point. Due to the local nature of the LIME method, which focuses only on the inputs and outputs of the model, it is possible to analyze the significance of the features of the input data for any of the black box models and hand over the results of the local feature significance analysis to a human for the next step[13]. Since the location of break in LOCA is a classification problem, LIME is used to analyze the importance of break location to monitoring parameters.

## 2.3 Overview of Diagnosis Work Flow

The aim of this paper is to achieve an aid to the operator in diagnosing accidents by developing an effective reactor accident diagnosis model and analyzing it by parameter significance. The workflow is therefore shown in Figure 4.

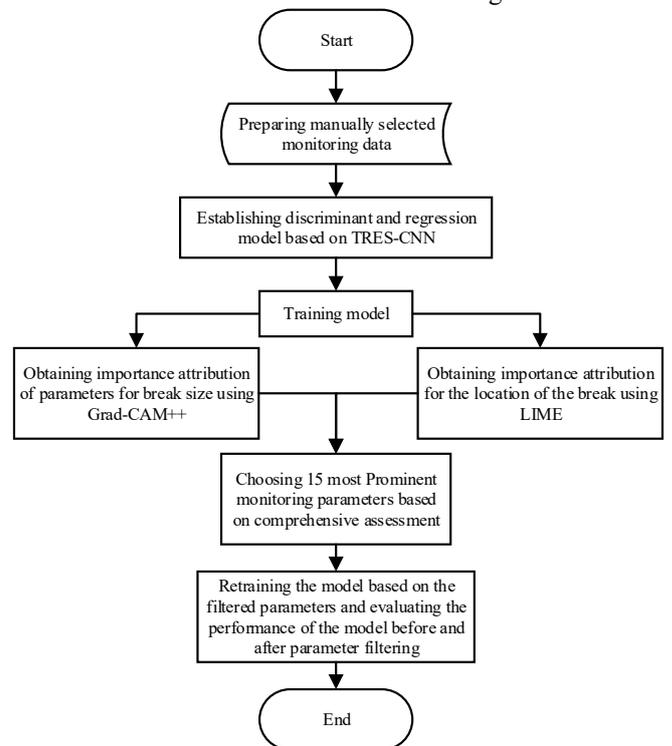

**Figure 4 Flow chart of monitoring parameter selection for LOCA diagnosis**

## 3. EXPERIMENT AND ANALYSES

### 3.1 Experimental Dataset

The accurate identification of the LOCA position and size is a primary consideration when LOCA occurs. Therefore, the information about the accident is provided for operators to take effective management.

To train a diagnostic model for LOCA, transient response data of high quality is required. Previous studies mostly attain transient data from full scale simulator or reactor transient analysis code, e.g., RELAP5, since the amount of transient data collected from real reactors is little and uneven. In this work, the transient data for model training is acquired from Advanced Reactor System Analysis Code (ARSAC), which is the transient analysis code developed by the Nuclear Power Institute of China (NPIC). Compared with other codes (e.g.,

CATHARE, WCOBRA/TRAC, RELAP5), ARSAC has the following characteristics: 1) Advanced matrix solution algorithm NRLU; 2) Refined wall heat transfer model; 3) Refined reflooding analysis module; 4) Advanced physical property analysis module[14]. In the modeling process, a full model of HPR1000 is constructed, and the schematic diagram of the RCS is roughly depicted in Figure 5.

In the process of data collecting, after 500s of steady state operation on ARSAC based on HPR1000 nuclear reactor, different simulated breaks are inserted, varying in diameter and position, e.g., at hot leg or cold leg. The initial 38 parameters are manually selected empirically, with time span of 100s and frequency of 2 samples per second. In total, 346 cases of LOCA transients varying in break sizes and positions are collected and randomly split into training set and testing set with 276 and 70 cases respectively.

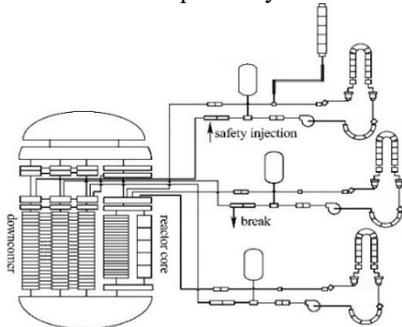

**Figure 5 Schematic diagram of HPR1000[15]**

### 3.2 Training Criterion

In the training process, the definition of loss function is crucial since it determines the behavior of the diagnostic model. The diagnosis has two goals, which are: 1) to predict the size of the break; 2) to distinguish whether it is a hot-leg break or a cold-leg break. As for the first goal, the loss function is defined by mean square error, as

$$\ell_{re}(x, y) = \text{sum}\{l_1, \ldots, l_N\}, \quad l_n = (x_n - y_n)^2 \quad (2)$$

where $x_n$ is the output break size of the diagnostic model, $y_n$ is the target value in the transient simulation and $N$ is the training minibatch dimension. When it comes to the second goal, considering it is a classification problem, the loss function is calculated by cross entropy loss function, as

$$\ell_{cl}(x, y) = \text{sum}\{l_1, \ldots, l_N\},$$
$$l_n = -\sum_{c=1}^{C} \log \frac{\exp(x_{n,c})}{\exp\left(\sum_{i=1}^{C} x_{n,i}\right)} y_{n,c} \quad (3)$$

where $x_n$ is the predicted probability distributions for categories, $y_n$ is the target one-hot encoding of the true transient type, $C$ is the number of classes, and $N$ is the minibatch dimension. Altogether, the loss function of the diagnostic model is defined as

$$\ell = \ell_{cl} + \ell_{re} \quad (4)$$

After defining the loss function, Adam optimizer is used as the stochastic gradient descent optimization method, which outperform other gradient descent optimizers in convergence and speed[16]. We set the learning rate $\alpha = 0.001$, two momentum parameters $\beta_1 = 0.9$ and $\beta_2 = 0.99$ respectively, and $\epsilon = 1 \times 10^{-8}$. Additionally, we have used dropout regularization to avoid over-fitting[17]. The training is terminated when the maximum comparable error within 10 iterations is less than 5%.

### 3.3 Evaluation of Diagnostic Model

Since there are two training targets in this paper, the categorial and numerical labels, respectively. Therefore, the Micro-F1 value and accuracy are used as the evaluation method for the category-based labels and the SSE is used as the performance metric for the numerical labels.

We compare our proposed methods TRES-CNN with the following baselines:
- BPNN: Basu and Barlett proposed a two-stage BP neural network-based method for reactor accident transient diagnosis[18];
- CNN-based: Lee et al. proposed a CNN-based reactor accident diagnosis model that transforms the raw data from the reactor monitoring system into a 2D image scheme with two channels[19]

Table 1 below illustrates how TRES-CNN performs in comparison to other baseline models. On the LOCA-based dataset constructed in this paper, TRES-CNN surpassed the baseline model by a significant margin in terms of prediction of break location and break size. In addition, the model performance is more significantly improved due to the introduction of residual block structure and dropout regularization in TRES-CNN.

**Table 1 Performance Comparison**

| Model | SSE Value (average) | Micro-F1 Value | Accuracy |
|---|---|---|---|
| BPNN | 0.543 | 0.882 | 0.905 |
| CNN-based | 0.404 | 0.944 | 0.949 |
| TRES-CNN (No residual-block or dropout) | 0.453 | 0.913 | 0.927 |
| TRES-CNN | 0.194 | 0.959 | 0.969 |

### 3.4 Parameter Significance and analysis

After training the TRES-CNN model, the location of LOCA can effectively be identified and the size of the break is accurately predicted. At this point, it can be considered that the model has learned the process of parameter change which is more significant for accident diagnosis, and this diagnostic knowledge is stored in the model through network parameters. Through the Post-hoc attribution analysis of the model, the local attention details of the input data can be obtained, and the knowledge learned by the model can be expressed in a way that human can understand. Since each input sample corresponds to a unique attention distribution, the attention distribution of all training data is linearly superimposed so that the lumped attention distribution of the diagnostic model to the training data can be learned. Such attention distribution is generally called saliency map, and the lumped saliency map for the input parameters of LOCA accident is shown in Figure 6.

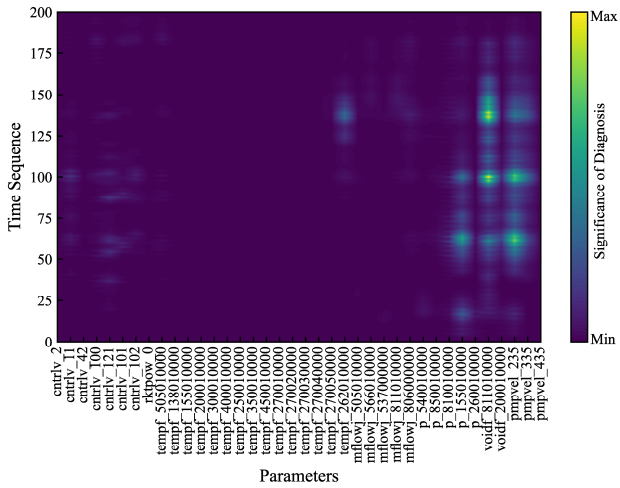

**Figure 6 Dumped saliency map of input dataset**

In Figure 6, the horizontal axis represents the point locations of several monitored parameters in the reactor's one-loop system. The abbreviations on the left side of the underline represent the names of the physical parameters, e.g. cntrlvar for the control variable defined by the monitoring system, tempf for the temperature of the liquid fluid within the control body, etc.; the numbers on the right side represent the different acquisition locations for that physical quantity, e.g. pmpvel_435 means the coolant pump speed of the third loop. The vertical axis is the time axis, and since the sampling time is 100s and the sampling frequency is 2 times per second, there are 200 valid samples for each parameter on the vertical axis. The lighter colour in the graph means that the "hotness" of the data is of more significance in that area, i.e. the model is more concerned with the change in values at those spaces. In order to calculate the overall significance of each parameter over the sampling time, the distributions of the data on the time axis were summed up and sorted to obtain a significance distribution for each parameter, as shown in Figure 7.

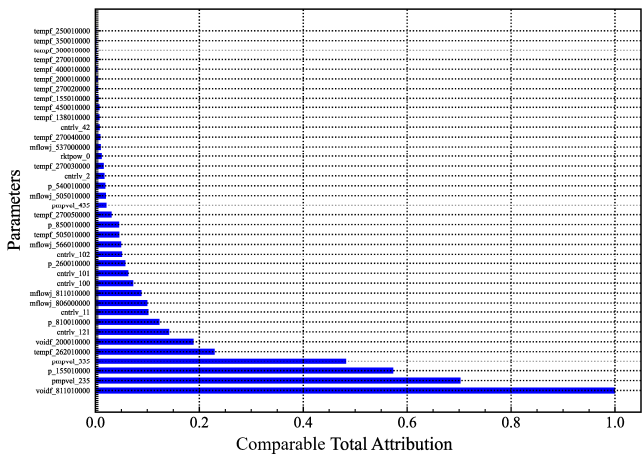

**Figure 7 Significance distribution of monitoring parameters**

It can be observed from Figure 7 that some of the parameters are able to play a more critical role in the diagnosis of incidents. Therefore, such significance distribution enables to obtain the parameters that are more critical for diagnosis of LOCA, in order to achieve effective dimensionality reduction of massive data in a dataset with more parameters. Next, outliers in the sample are removed, and the values that were more than 1.5 times interquartile spacing away from the range of the 1st and 3rd quartiles are removes as outliers[20]. Thus, the selected top 15 significant parameters with their sample distributions are depicted in the violin plot as Figure 8.

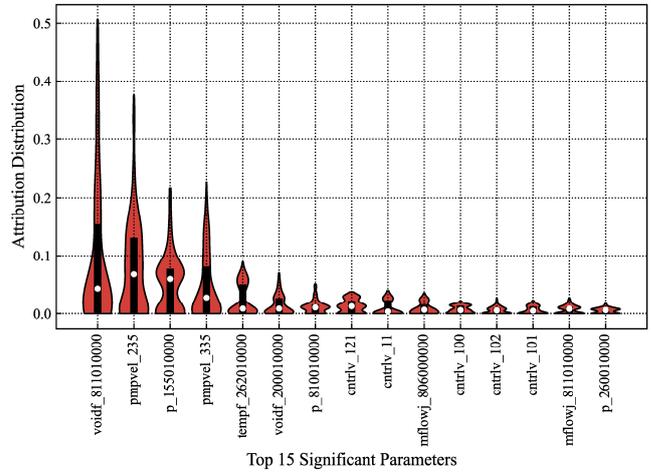

**Figure 8 Top 15 significant parameters with sample distributions**

The width of each violin part represents the density of distribution. The two ends of the black line in the middle denote the upper and lower 1/4 quartiles of the data, while the white circle dot in the middle stands for the median value of each data group. Among the 15 most significant parameters for diagnosing LOCA, the top three are voidf_811010000, pmpvel_235 and p_155010000. The physical meanings of these parameters are the safety injection tank level, the breakout loop coolant pumping rate and the reactor coolant pressure.

In the violin diagram, there is a clear difference in the distribution pattern of each data component. The distribution of some of the parameters is characterized by distinct spikes, which represent a greater differentiation in the contribution of that parameter to the diagnosis. For example, the parameter with the most significant overall influence on the diagnosis, i.e. the safety injection tank level, has most of its data concentrated near the bottom of the x-axis, but some of the data has a very high significance score, resulting in the parameter being ranked first in terms of overall significance score. This is due to the fact that at smaller breach sizes, the injection of the safety injection tank is not triggered within the monitored time frame and thus does not provide a contribution to the diagnosis of breach location and size.

From a physical point of view, parameters selected by parameter significance analysis based on interpretable deep learning do have more valid diagnostic implications. As parameter pump_235, namely the change of coolant pump velocity in the first loop (break loop) of the reactor, is a characteristic parameter that can well reflect the LOCA diagnosis results. In principle, the decrease in the pump speed of the main pump is caused by the sharp increase in the local void fraction. When the break is located in the cold leg, since it is closer to the position of the coolant pump than the outlet section of the pressure vessel, the void fraction at the pump will increase in a very short period of time, resulting in an earlier decline in the pump speed. In addition, since the size of the break has a significant indigenous effect on the increase rate of the void fraction, the drop time and speed of the main pump in the break loop are effective basis for the reverse deduction of the size of the break.

### 3.5 Using Selected Parameters to Retrain TRES-CNN

Selecting parameters that are more meaningful in the accident diagnosis process can, on the one hand, significantly reduce the number of parameters that the operator needs to monitor to aid manual diagnosis; on the other hand, it can significantly reduce the amount of data for model input when training a deep learning-based accident diagnosis model, ensuring that the model can learn valid information for diagnosis from valid data, while avoiding overfitting of the model during training. Therefore, it is necessary to compare the models trained before and after data filtering.

When evaluating the impact of the choice of parameters on the model, it is important to assess not only the impact of changes in the amount of data on the model performance, but also the impact on the model training process. Micro-F1 value, accuracy, and SSE are therefore chosen as indicators of model performance, and the gradient descent of the loss function using the Adam optimizer during model training is compared.

**Table 2 Different Input Parameters' Impacts on Diagnostic Model Performance**

| Input data | SSE | Micro-F1 value | Accuracy |
|---|---|---|---|
| Total 38 parameters | 0.194 | 0.959 | 0.969 |
| Top 15 significant parameters | 0.203 | 0.954 | 0.959 |
| Relative error | +4.64% | -0.52% | -1.03% |

As can be seen in Table 2, in the models before and after the extraction of the effective parameters, although there is a slight decrease in model performance due to the reduction in the number of parameters, the relative change is only 0.521% and 1.032% for both Micro-F1 and Accuracy respectively.

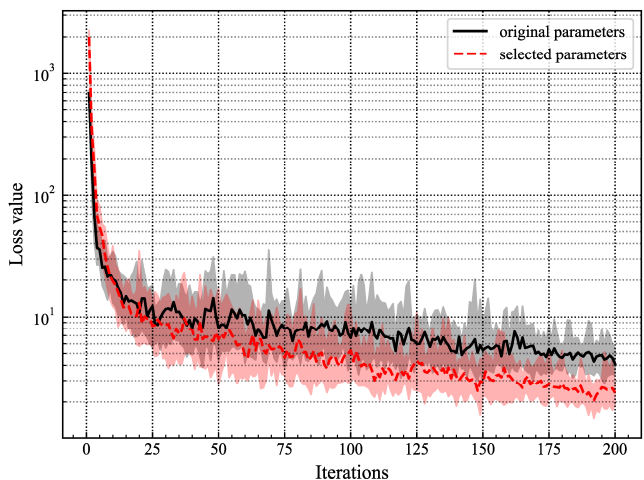

**Figure 9 Loss reduction during initial training and retraining process**

During the training of the model, the iteration speed of the loss function has been significantly improved due to the reduction in the number of parameters. As depicted in Figure 9, whereas training with 38 monitoring parameters as input required 200 iterations, training with the filtered input parameters requires only about 113 iterations to converge.

## 4. CONCLUSION

In this paper, we first propose a novel deep learning-based diagnostic model for reactor accident diagnosis, which not only takes into account the time-series characteristics of the data, but also improves the overfitting of the training process through residual concatenation and dropout mechanisms. Secondly, a hybrid interpretability approach based on Grad-CAM++ and LIME is used to analyze the validity of the pre-selected manually selected input parameters on the diagnostic results, and the most important parameters of concern are extracted as parameter inputs to the reduced dimensional diagnostic model. Thus, on the one hand, making the training of the model much faster, and on the other hand providing fewer but more effective parameters to assist the operator in manual diagnosis.